\documentclass[fleqn,usenatbib]{mnras}

\usepackage{newtxtext,newtxmath}

\usepackage[T1]{fontenc}

\DeclareRobustCommand{\VAN}[3]{#2}
\let\VANthebibliography\thebibliography
\def\thebibliography{\DeclareRobustCommand{\VAN}[3]{##3}\VANthebibliography}

\usepackage{graphicx}	
\usepackage{amsmath}	
\usepackage{subcaption}

\newcommand{\changeone}[1]{#1}
\newcommand{\changetwo}[1]{#1}

\title[Topology of galaxies]{Topological data analysis reveals differences between simulated galaxies and dark matter haloes}

\author[A. Ouellette et al.]{
Aaron Ouellette,$^{1,2}$\thanks{E-mail: aaronjo2@illinois.edu}
Gilbert Holder,$^{1}$ Ely Kerman$^{3}$
\\
$^{1}$Department of Physics, University of Illinois at Urbana-Champaign, Urbana, IL 61801, USA \\
$^{2}$Center for AstroPhysical Surveys, National Center for Supercomputing Applications, Urbana, IL 61801, USA \\
$^{3}$Department of Mathematics, University of Illinois at Urbana-Champaign, Urbana, IL 61801, USA
}

\date{Accepted XXX. Received YYY; in original form ZZZ}

\pubyear{2023}

\begin{document}
\label{firstpage}
\pagerange{\pageref{firstpage}--\pageref{lastpage}}
\maketitle

\begin{abstract}
We use topological summaries based on Betti curves to characterize the large-scale spatial distribution of simulated dark matter haloes and galaxies. Using the IllustrisTNG and CAMELS-SAM simulations, we show that the topology of the galaxy distribution is significantly different from the topology of the dark matter halo distribution. Further, there are significant differences between the distributions of star-forming and quiescent galaxies. These topological differences are broadly consistent across all simulations, while at the same time there are noticeable differences when comparing between different models. Finally, using the CAMELS-SAM simulations, we show that the topology of the quiescent galaxies in particular depends strongly on the amount of supernova feedback. These results suggest that topological summary statistics could be used to help better understand the processes of galaxy formation and evolution.
\end{abstract}




\section{Introduction}\label{sec:intro}

One of the biggest challenges in cosmology today is understanding the effects of galaxy formation and evolution on the distribution of matter in the universe \citep{Daalen2011,Chisari2019}. Many upcoming observational programs, such as DESI \citep{DESICollaboration2016}, LSST \citep{Ivezic2019}, Euclid \citep{Laureijs2011}, CMB-S4 \citep{Abazajian2019}, and others, are poised to collect vast amounts of data that will help further constrain cosmological models, but robust models of baryonic effects are necessary in order to ensure that statistical uncertainties are not overwhelmed by systematic ones.

On the theory side, large cosmological simulations aim to simulate the formation and evolution of galaxies and the growth of large-scale structure from first principles. It is now well established through cosmological hydrodynamic simulations such as EAGLE \citep{Schaye2015,Crain2015}, Illustris \citep{Nelson2015,Vogelsberger2014b}, and IllustrisTNG \citep{Nelson2019} that feedback from active galactic nucleii (AGN) and supernovae is vital to produce realistic galaxies. However, these simulations all have very different implementations of feedback models, leaving some uncertainty in their specific predictions. Consequently, it is necessary to develop statistical probes that are best suited to constrain these models.

The usual statistical probe of large-scale structure is the 2-point correlation function or its Fourier transform, the power spectrum. This is the ideal summary statistic in the case of Gaussian random fields, which are fully described by their 2-point functions, but cosmological  density fields are  non-Gaussian. An increasing amount of literature suggests that probes that encode information about the higher-order moments of clustering beyond the 2-point correlation function are extremely useful in providing cosmological constraints. In addition to the higher $n$-point functions, examples of these probes include: counts-in-cells \citep{Peebles1980,Leicht2019,Uhlemann2020,Perez2022}, void probability functions \citep{White1979,Conroy2005,Perez2021}, nearest neighbor distributions \citep{Banerjee2021}, network statistics \citep{Coutinho2016,Hong2016,Naidoo2020}, and many others. These probes generally have the advantage that they are relatively easy to compute, and they include information from higher-order clustering. On the other hand, they are usually much harder to predict directly from theory compared to the power spectrum or 2-point function. 

It is now well known that on large scales, the distribution of matter in the universe forms a highly complex and connected structure known as the cosmic web \citep{Bond1996,Weygaert2008,Libeskind2018}. This structure naturally lends itself to description through the ideas of topology. The topology of the cosmic web has been studied since the advent of large redshift surveys \citep{Gott1986,Hamilton1986,Weinberg1987,Gott1987,Gott1989,Melott1988,Melott1989}. Recent advances in the fields of computational topology and topological data analysis (TDA) have provided principled and efficient methods for statistically describing the "shape of data." This is often done using a formalism called \changeone{persistent homology \citep{Edelsbrunner2002,Edelsbrunner2010} where one characterizes the topological features present in data as a function of scale. An overview of these methods can be found in \cite{Carlsson2009}}.

These methods from TDA have just recently started to see applications in cosmology and astrophysics, but the breadth of applications has been quite wide. \cite{Weygaert2010,Weygaert2011} proposed using persistent homology and Betti numbers to characterize the large-scale structure of the cosmic web and \cite{Sousbie2011} used ideas from topology to develop a robust method of identifying clusters, filaments, and voids in the cosmic web. Studies on dark matter (DM) haloes and the cosmic web were done by \cite{Pranav2017,Bermejo2022,Tsizh2023}. Other applications include \cite{Xu2019}, who developed a formalism to identify statistically significant voids and filaments. \cite{Kono2020} used persistent homology to identify the baryon acoustic oscillation features in the spatial distribution of galaxies. \cite{CisewskiKehe2022} showed that TDA is able to discriminate between different DM models using only subhalo spatial distributions. Persistent homology has also been applied studies of the period of re-ionization \citep{Elbers2019,Elbers2023,Thelie2022}, the analysis of weak lensing data \citep{Heydenreich2021,Heydenreich2022}, \changeone{analysis of the cosmic microwave background \citep{Pranav2019b}}, and studies of non-Gaussianity in the primordial density fluctuations \cite{Cole2020,Biagetti2021,Biagetti2022}. 
Most similar to our work is that of \cite{Bermejo2022}, where the authors explored the topology of DM haloes using persistent homology. They reported a ``topological bias": haloes exhibited different topologies dependent on their mass. 

In this work, we use Betti curves from persistent homology to characterize the large-scale structure of galaxies and DM haloes,  taking into account the potentially confounding effects of number density. We then show that these topological summaries reveal differences in the large-scale structure of haloes and different galaxy populations. Our main results are as follows. First, we \changeone{calculate Betti curves for DM haloes and show that they are sensitive to the higher moments of clustering beyond the 2-point correlation function. The Betti curves for galaxies are very sensitive to small scale features.} We find that a second physical scale shows up in the topology of galaxies that is not present in the haloes. Further, separating galaxies into star-forming and quiescent sub-samples reveals that the topology of quiescent galaxies is especially sensitive to differences in the feedback model used. Based on these results, we argue that galaxies are topologically distinct from DM haloes and that topological summaries could provide very useful information in order to better constrain models of galaxy formation and evolution.

This paper is structured as follows. In \autoref{sec:tda} we review the basics of topological data analysis. Our analysis pipeline is outlined in \autoref{sec:methods}. Next we calibrate our method in \autoref{sec:rand} by exploring the topology of samples of random points. We introduce the simulations and mock galaxy catalogs used in this work in \autoref{sec:sims}. In \autoref{sec:results} we present our main results.  We conclude in \autoref{sec:conlusion}.

\section{Overview of Topological Data Analysis}\label{sec:tda}
Topology in the traditional sense is concerned with properties of spaces that are invariant under continuous deformations. One of the most intuitively familiar formulations of topology is \textit{homology} -- the characterization of ``holes" in a topological space. Formally, this is described through homology groups, the rank of which represents the number of independent topological features. The full formal mathematical description of homology is beyond the scope of this work, but is laid out in many textbooks on algebraic topology, such as \cite{Hatcher2001}. More details on the computational aspects of topology can be found in \cite{Edelsbrunner2010}. Here, we will focus on providing an intuitive summary that will be useful for understanding applications of topology to cosmological datasets.

\changeone{An orientable topological space of dimension $d$ has $d+1$ possible types of holes. The classes of holes of type $j$ form a group, $H_j$, whose rank is the $j^{th}$ Betti number of the space, $\beta_j$. For example, $\beta_0$ is the number of connected components, $\beta_1$ is the number of independent classes of non-contractible loops, and $\beta_d$ is the number of compact connected components. As a simple example, the two-sphere has Betti numbers $\beta_0=1$, $\beta_1=0$, $\beta_2=1$ (one connected component, zero non-contractible loops, and one compact connected component), while the two-torus has Betti numbers $\beta_0=1$, $\beta_1=2$, $\beta_2=1$ (in the torus there are two independent classes of non-contractible loops). In a cosmological setting, we are primarily interested in the homology groups $H_0$, $H_1$, and $H_2$. These can be thought of as (very loosely) corresponding to clusters, filaments, and voids in the cosmic web.}

\subsection{Persistent homology}
Persistent homology \citep{Edelsbrunner2002,Carlsson2005,Ghrist2007} \changeone{is a formalism to study the homology of a topological space as a function of a single parameter, such as scale}. This turns out to be ideal in a cosmological setting since the properties of the matter distribution are very much scale dependent: at very large scales the universe looks homogeneous and isotropic, at intermediate scales we get the filamentary cosmic web, and at small scales we might see individual clusters of galaxies.

Here, we focus on the application of persistent homology to data in the form of a point cloud\footnote{Persistent homology can also be applied to data in the form of scalar fields, such as density fields, as is done in \cite{Pranav2019,Pranav2021}.} (e.g., galaxy catalogs). In order to compute topological quantities, we need some way of transforming the disconnected points into a topological space that in some way captures the ``shape" of the underlying data. This is often done using \textit{simplicial complexes} which can be thought of as a triangulation of the topological space represented by the point set. Formally, a $k$-simplex is the convex hull of $k+1$ affinely independent points. \changeone{The boundary of a $k$-simplex consists of smaller simplices called its faces. For example, the faces of a 3-simplex (a tetrahedron) consist of four 2-simplices (triangles), six 1-simplices (edges), and four 0-simplices (vertices). A collection of simplices $K$ is a simplicial complex if each face of each simplex of $K$ is also in $K$, and any two simplices of $K$ which intersect do so along a common face. Simplicial complexes yield simple combinatorial descriptions of topological spaces and provide a standard way of computing homology groups.}

Simplicial complexes can be associated to a point cloud at different scales, leading to a nested family of complexes called a \textit{filtration}. One of the commonly used filtrations is the Vietoris-Rips filtration whose scaling parameter is defined in terms of the pairwise distances between points in the dataset. The simplicial complex of the Vietoris-Rips filtration at scale $r$ is comprised of the points in the given point cloud together with all the higher dimensional simplices whose vertices are all within a distance $r$ from one another.

For large datasets, Vietoris-Rips filtrations become prohibitively expensive to compute, since calculating pairwise distances scales quadratically with the number of points. For this reason, we use a filtration based on alpha complexes \citep{Edelsbrunner1983,Edelsbrunner1994}, which are subcomplexes of the Delaunay triangulation. \changeone{The filtration value $\alpha^2$ of a simplex of the Delaunay triangulation is defined to be the square of its circumradius if the interior of the corresponding circumsphere does not contain any vertices of the Delaunay triangulation. Otherwise, if there is a vertex in the interior, the filtration value of the simplex is defined to be the minimum filtration value of the codimension one faces contained in the interior \citep{Rouvreau2022}. \footnote{Specific implementations of the alpha complex differ on whether they use $\alpha$ or $\alpha^2$ as the filtration parameter. Regardless, we use $\alpha$ throughout this paper which has units of distance.} The alpha complex for $\alpha \geq 0$ is then defined to be the subcomplex of the Delaunay triangulation consisting of simplices with filtration value at most $\alpha^2$.} The construction cost of the alpha complex scales sub-quadratically with the number of points and exponentially with the number of dimensions, making the alpha complex more efficient than the Rips complex for large numbers of points in low dimensions.

\begin{figure}
    \centering
    \begin{subfigure}{\linewidth}
        \centering
        \includegraphics[width=\linewidth]{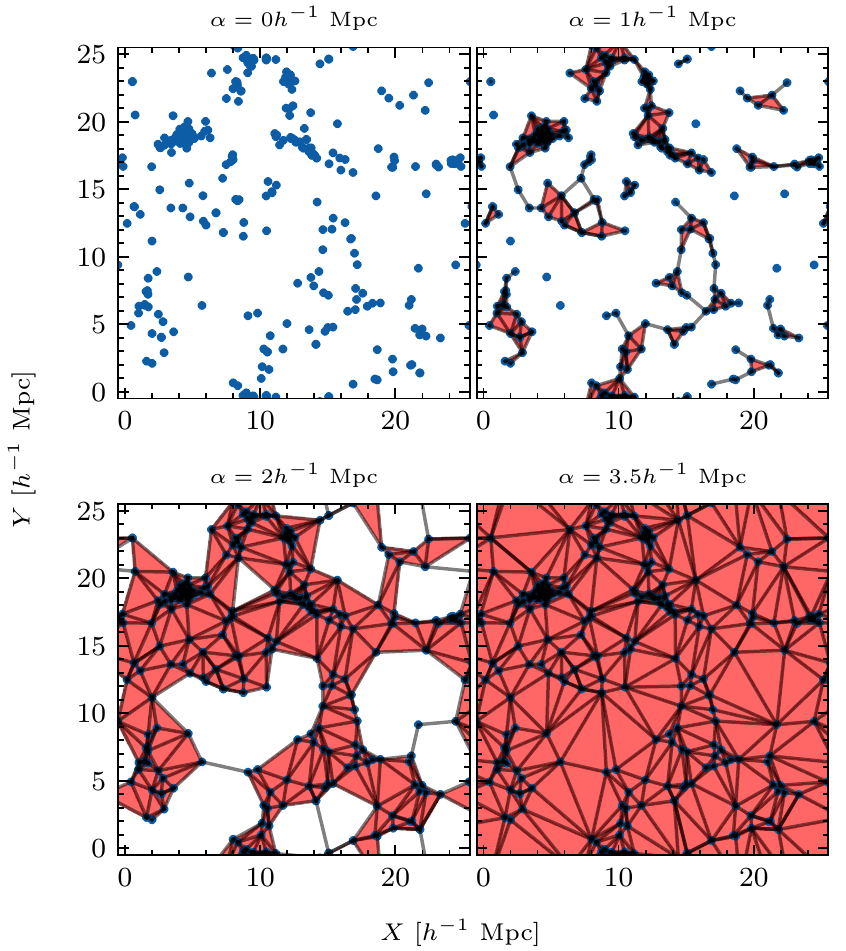}
    \end{subfigure}
    \hfill
    \begin{subfigure}{0.75\linewidth}
        \centering
        \includegraphics[width=\linewidth]{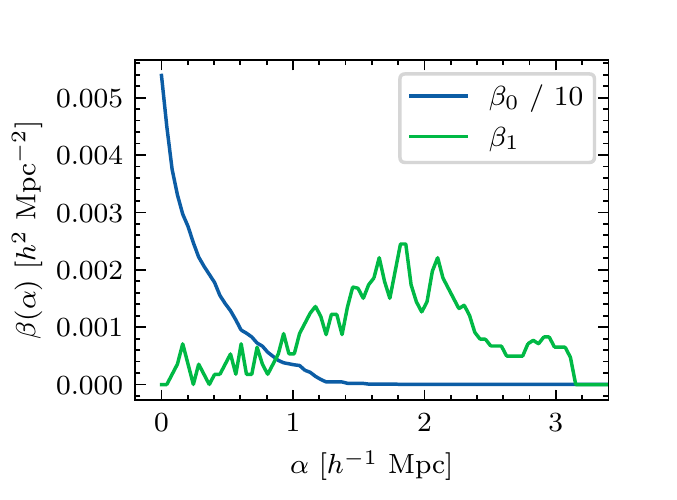}
    \end{subfigure}
    \hfill
    \caption{\textit{Top}: Two-dimensional visualization of the evolution of the alpha complex for a slice of galaxies. Here we took galaxies in a 6 Mpc slice from the EX0 simulation of the CAMELS IllustrisTNG suite. The panels show the resulting alpha complexes for increasing values of $\alpha$. \textit{Bottom}: Betti curves, showing the full evolution of the number of homology features as a function of scale. Note that the Betti curves are given per unit area since this example is set in 2D.}
    \label{fig:example}
\end{figure}

Once we have chosen a filtration, it is fairly straightforward to compute the homology of the resulting simplicial complexes as a function of scale (represented by the filtration parameter). As the filtration parameter $\alpha$ grows, a homology class may appear at some value $\alpha_{\text{birth}}$, persist for some range, and then may disappear at some larger value $\alpha_{\text{death}}$ (this will be illustrated in more detail in \autoref{fig:example}, see \autoref{sec:methods}). These pairs of births and deaths are tracked in a \textit{persistence diagram} in which each is represented by its corresponding interval, $[\alpha_{\text{birth}}, \,\alpha_{\text{death}}]$. The persistence diagram will also contain one infinitely long interval of the form $[\alpha_{\text{birth}}, \,\infty)$ for each nontrivial homology class of the ambient space in which the data lives. \changeone{These are formally known as the essential homology classes of the manifold.} For example, in Euclidean space, there will be a single infinite interval in the $H_0$ perisitence diagram, representing the fact that above a certain scale all simplices will become connected.

\subsection{Topological summaries}
Persistence diagrams provide useful summaries of the topological features in a dataset, but they are not convenient to work with statistically. While measures of differences between two diagrams exist (e.g. the bottleneck distance or $p$-Wasserstein distance), they are very computationally intensive since some optimal matching between points in different diagrams is necessary. Additionally, a mean can be defined on the space of persistence diagrams \citep{Turner2014}, but it is not necessarily unique and suffers from the same computational challenges.

The standard way of solving this problem is to define a functional summary of persistence diagrams \citep{Berry2020}. This is a mapping from the space of persistence diagrams to the space of functions of the filtration parameter that encodes in some way the information present in a persistence diagram. There are many ways to do this that have been defined in the TDA literature, some common examples are: Betti curves, landscape functions \citep{Bubenik2012}, weighted silhouettes \citep{Chazal2013}, and entropy summary functions \citep{Atienza2020}. Each of these functional summaries has different tradeoffs between interpretabillity, robustness to noise, and the amount of information retained from the persistence diagram. \changeone{We also note that \cite{Adler2017} have introduced a way to directly model persistence diagrams as random point processes, but standard functional summaries are still significantly cheaper to compute.}

In this work, we focus on Betti curves, since they are perhaps the simplest and most interpretable summary available. A Betti curve is a direct generalization of Betti numbers: for a given homology dimension $d$, $\beta_d(\alpha)$ is simply the $d^{\text{th}}$ Betti number of the simplicial complex at the corresponding value of the filtration parameter.

\section{Analysis pipeline}\label{sec:methods}
In general, the analysis pipeline to apply TDA to point cloud data consists of the following steps.
\begin{enumerate}
    \item Define a simplicial complex on the point cloud that depends on a filtration parameter.
    \item For each value of the filtration parameter, calculate the homology of the resulting complex.
    \item Track the birth and death scales of the homology features in a persistence diagram.
    \item Calculate a functional summary of the persistence diagram in order to statistically quantify the topology.
\end{enumerate}

In this work, the data consists of galaxy (or halo) positions. These galaxies and haloes come from cosmological simulations that use periodic boundary conditions, hence our data and alpha complexes naturally live in the three dimensional torus instead of $\mathbb{R}^3$. We exclude the topology of the ambient torus, leaving only the topology of the galaxy and halo distributions, by deleting all infinitely long intervals in our persistence diagrams. The resulting persistence diagrams are then used to compute Betti curves that track the number of homology features as a function of scale. We report Betti numbers normalized by volume, i.e., the number density of various homology features.

We use a Python wrapper\footnote{\url{https://github.com/ajouellette/alpha-complex-wrapper}} around the C++ library \verb|GUDHI|\footnote{\url{https://gudhi.inria.fr}} \citep{GUDHIProject2022} to construct periodic alpha complexes and compute their persistent homology. Betti curves and other summary functions are implemented in the \verb|representations| module of the \verb|GUDHI| Python package \citep{Dlotko2022}.

To provide a concrete example of this process, in \autoref{fig:example} we show how an alpha complex filtration is constructed for a small set of galaxies from a cosmological simulation. We selected galaxies in a $6h^{-1}$ Mpc thick slice from the EX0 simulation of the CAMELS IllustrisTNG suite\footnote{\url{https://camels.readthedocs.io/en/latest/}}. For visualization purposes, this example is set in 2D. The top 4 panels in \autoref{fig:example} show the resulting alpha complexes for given values of the filtration parameter $\alpha$. As we increase $\alpha$, the number of disconnected components (or separate clusters of points) decreases rapidly while loops gradually form and then disappear at larger values of $\alpha$. Finally, at a scale of $3.5h^{-1}$ Mpc, only one large connected component remains. If we track this evolution for all values of $\alpha$, we can compute the Betti curves (normalized by the area of the 2D box) for homology dimensions 0 and 1. These are shown in the bottom panel of \autoref{fig:example}, where we see the rapid decrease in the number of disconnected components ($\beta_0$) and a rise and then fall in the number of loops ($\beta_1$).

\section{Topology of random points}\label{sec:rand}
In order to rigorously compare the topologies of different galaxy samples, we first explore how the topology depends on the properties of various random point processes. Specifically, we look at uniformly distributed points and points sampled from log-normal fields. In both cases, the points are sampled inside a 3D periodic box of size $L$.

\begin{figure}
    \centering
    \includegraphics[width=\linewidth]{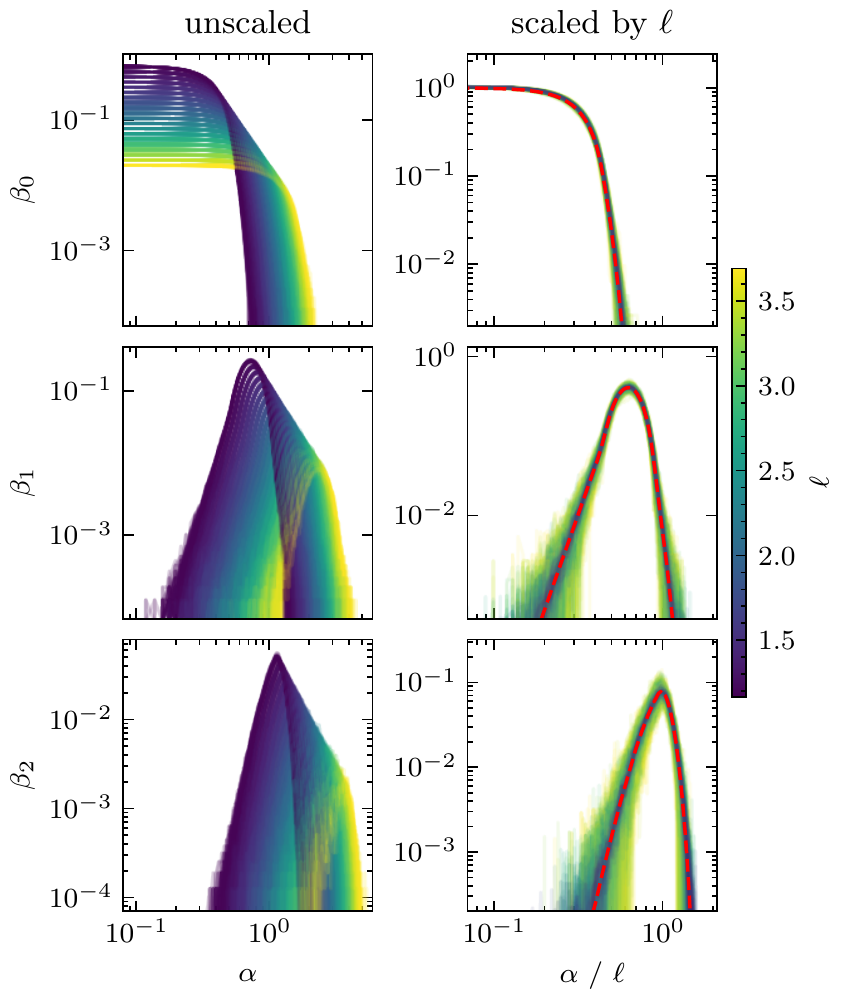}
    \caption{Betti curves for Poisson point processes. \textit{Left column}: 0-, 1-, and 2-dim Betti curves for 1,000 different samples of uniformly distributed points with $L=25$. The colors indicate the mean particle separation $\ell = L / N^{1/3}$ for each sample of random points. $\alpha$ has arbitrary units of length and the Betti curves have arbitrary units of inverse volume. \textit{Right column}: scaled Betti curves for the same 1,000 random point samples. The dashed red line indicates the average scaled Betti curve for a larger box with $L=75$.}
    \label{fig:poisson_betti}
\end{figure}

\begin{figure}
    \centering
    \includegraphics[width=2.4in]{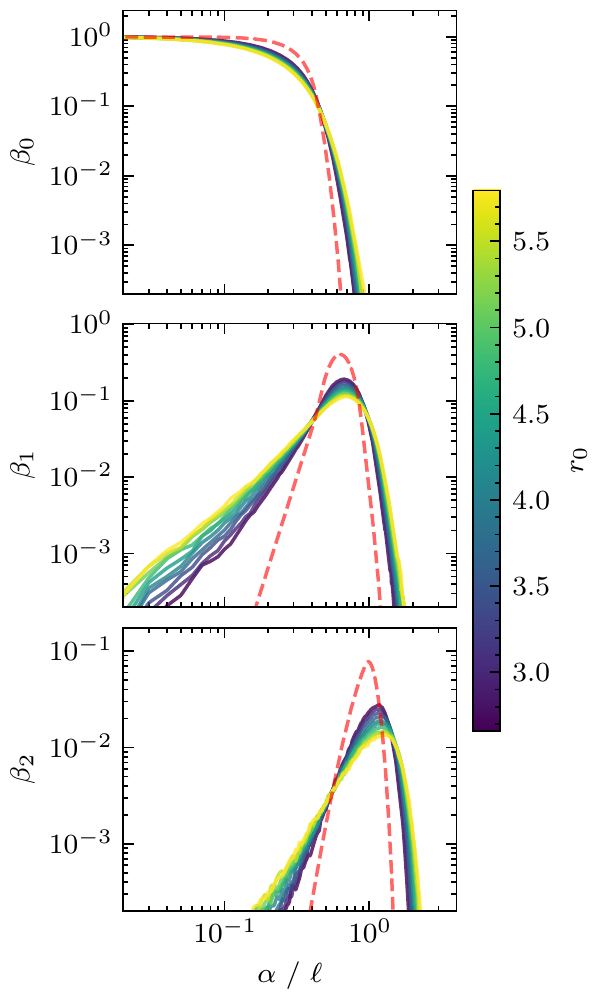}
    \caption{Scaled Betti curves for log-normal point processes following a power-law correlation function. The mean Betti curves for a Poisson point process are overlaid with the red dashed line.}
    \label{fig:lognormal_betti}
\end{figure}

\subsection{Uniformly distributed points}
First, we consider the simplest case -- random points uniformly distributed inside a cube with sides $[0, L)$ (i.e., a Poisson point process). In this case, the points are characterized only by the average number density $n = N/L^3$ or mean separation $\ell = 1/n^{1/3}$. Since the persistent homology of a point cloud directly depends on the pairwise distances between the points, we expect the topological summaries to be strongly dependent on $\ell$, especially in the case where this is the only characteristic length of the point sample.

To test this, we generated 1,000 samples of random points uniformly distributed in a 3D periodic box of size $L = 25$ with a wide range of values for $\ell$. For each collection of points, we computed their persistent homology using periodic alpha complexes and calculated Betti curves for each of the 3 homology dimensions. The results are shown in in the left column of \autoref{fig:poisson_betti}. We see a strong dependence on $\ell$, however the curves for a given dimension generally have the same shape.

We would like a topological summary that does not depend on the number density of points, but rather captures the intrinsic shape of the data. In this very simple case of uniformly distributed points, the solution is very simple. All point samples should become self-similar if we scale all distances by $\ell$. Applying this to the Betti curves, we scale the filtration parameter by $\ell$ and re-normalize the Betti curves by a volume of $\ell^3$. The results are shown in the right column of \autoref{fig:poisson_betti}. We see that the scaled curves are now roughly identical; the outliers tend to be the point samples with the fewest points (corresponding to larger statistical noise). These curves now show a summary of the universal topology of samples from Poisson point processes. Additionally, we confirm that this topological summary does not depend significantly on the box size. We applied the same method to 1,000 realizations inside a larger box with $L=75$ and calculated the average Betti curve for each homology dimension. These curves are shown in the same figure with the red dashed lines and match the previous curves very well.

We note that similar results can be found in the math literature in relation to percolation theory or the theory of random graphs \citep[for example]{Robins2002,Robins2006}. \changeone{Our results also agree with the recent study by \cite{Gluzberg2023}: our scaled Betti curves correspond to their dimensionless quantity $\beta_{3,d}$ defined in their proposition 3.2.}

\subsection{Log-Normal fields}
Galaxy samples are not well represented by uniformly distributed random points. Next, we explore how clustering affects the topological summary functions.

To mimic the clustering of galaxies, we generate log-normal fields \cite[suggested as a model for galaxy distributions by][]{Coles1991} with a power-law power spectrum. In real space, this results in a 2-point correlation function that can be written as
\begin{equation}
    \label{eq:xi}
    \xi(r) = \left(\frac{r}{r_0}\right)^\gamma.
\end{equation}

We use the Python package \verb|powerbox|\footnote{\url{https://github.com/steven-murray/powerbox}} \citep{Murray2018} to generate 500 samples from log-normal fields with power-law clustering using a fixed exponent of $\gamma = -1.5 \pm 0.1$ and a wide range of values for the clustering length $r_0$. As before, for each sample we calculate its persistent homology and convert the persistence diagrams into scaled Betti curves. To reduce statistical noise, we also bin the samples by their clustering length and average the corresponding Betti curves. 

In \autoref{fig:lognormal_betti}, we plot the averaged scaled Betti curves for 8 bins between $r_0 = 3$ and $r_0 = 6$, with the mean scaled Betti curves for a Poisson point process overlaid. We see that, compared to the number density, the clustering of points is a small effect, causing slight shifts in the Betti curves. Increasing the clustering length suppresses the peak of the $H_1$ and $H_2$ Betti curves and lengthens the tails.

\section{Cosmological Simulations}\label{sec:sims}
We use galaxy and halo catalogs derived from two simulation suites: IllustrisTNG\footnote{\url{https://www.tng-project.org/}} and CAMELS-SAM\footnote{\url{https://camels-sam.readthedocs.io/en/main/}}. We chose these simulations in order to compare roughly similar models at a large enough size (box sizes of at least $\sim100h^{-1}$ Mpc) so that the results are less affected by sample variance.

The IllustrisTNG simulations \citep{Springel2018,Nelson2018,Pillepich2018,Marinacci2018,Naiman2018} are large high-resolution cosmological magnetohydrodynamic simulations that aim to fully model the galaxy formation process in a cosmological context. Here we use publicly released data from the TNG300 run \citep{Nelson2019}. 
The TNG simulations were run using the AREPO code \citep{Weinberger2020}. TNG300 uses a simulation box of comoving size $L = 205\, h^{-1}$Mpc and tracks the evolution of $2\times2500^3$ DM particles and gas cells. The simulation assumes a standard flat $\Lambda$CDM universe with $\Omega_m = 0.3089$, $\Omega_b = 0.0486$, $\sigma_8 = 0.8159$, $n_s = 0.9667$, and $h = 0.6774$.
Additionally, a complex supernova and AGN feedback model is used to model realistic galaxy formation \citep{Weinberger2017,Pillepich2018}. The halo and galaxy catalogs for the TNG simulations were generated by running \verb|SUBFIND| \citep{Springel2001}.

The CAMELS-SAM simulation set is a large suite of simulated galaxies that specifically aims to test the effects of wide ranges of both cosmological and astrophysical parameters \citep{Perez2022,VillaescusaNavarro2022}. CAMELS-SAM is based on 1,005 DM-only simulations inside a box with $L = 100h^{-1}$ Mpc, tracking the evolution of $N = 640^3$ particles using AREPO. These simulations cover a broad space of cosmological models by varying $\Omega_m$ (between 0.1 and 0.5) and $\sigma_8$ (between 0.6 and 1.0), but otherwise assume a standard flat $\Lambda$CDM model with $\Omega_b = 0.049$, $h=0.6711$, and $n_s = 0.9624$. Halo catalogs and merger trees were generated using \verb|Rockstar| \citep{Behroozi2013b} and \verb|ConsistentTrees| \citep{Behroozi2013}. Galaxy catalogs were generated by running multiple iterations of the Santa Cruz semi-analytic (SC-SAM) model of galaxy formation \citep{Somerville1999,Somerville2008,Somerville2015}, varying the amount of supernova and AGN feedback. The feedback in the SC-SAM model is varied in these simulations through three normalization parameters: $A_\text{SN1}$, $A_\text{SN2}$, and $A_\text{AGN}$.

The CAMELS-SAM simulations are further divided into 3 subsets: CV (cosmic variance), 1P (one parameter), and LH (latin hypercube). The CV set consists of 5 simulations using the fiducial model ($\Omega_m = 0.3$, $\sigma_8 = 0.8$, $A_\text{SN1} = 1$, $A_\text{SN2} = 0$, and $A_\text{AGN} = 1$) but with random initial seeds in order to quantify the amount of sample variance. The 1P set consists of 12 simulations that vary the three astrophysical parameters one at a time, keeping all other parameters at their fiducial values. Finally, the LH set consists of 1,000 simulations that simultaneously vary all 5 parameters and the initial seed.

Ideally, we would have also used the full hydrodynamic simulations from the CAMELS suite, but the box sizes of $25h^{-1}$ Mpc were too small for our purposes. Each simulation box has on average only $\sim800$ galaxies, making sub-selections based on various properties not viable. Additionally, sample variance due to the small box size overwhelmed variances due to parameter variations.

\begin{figure}
    \centering
    \includegraphics[width=\linewidth]{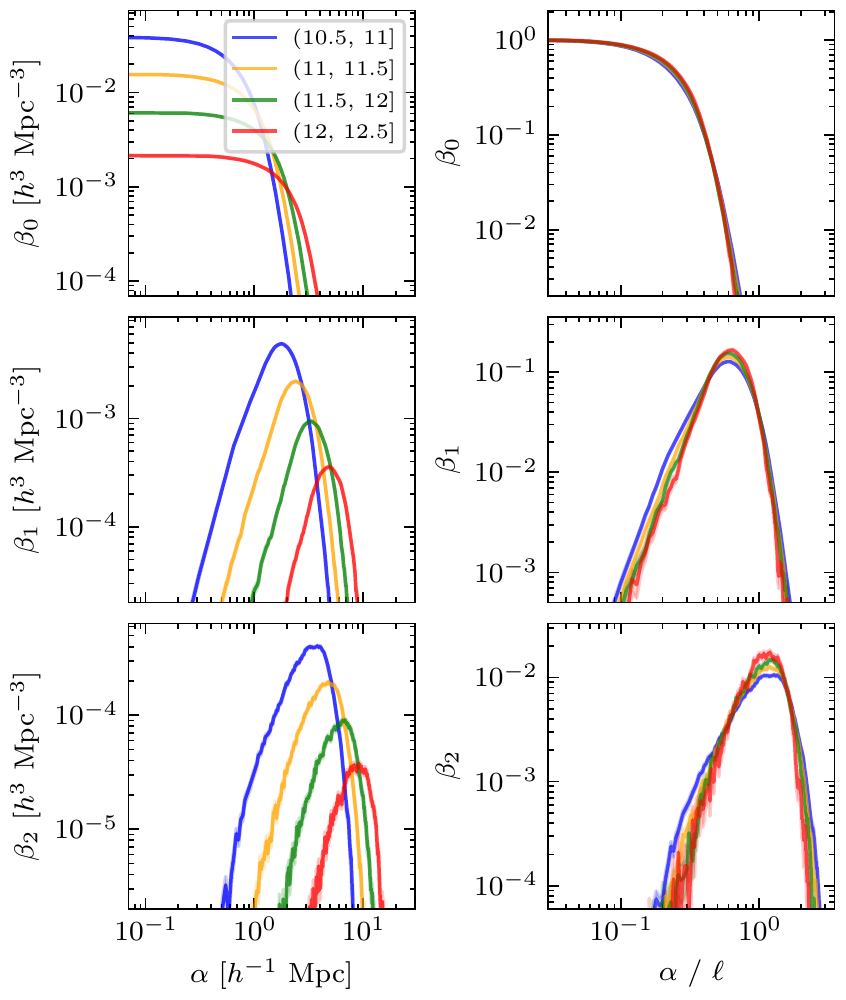}
    \caption{Betti curves for TNG300 haloes in four different mass bins. $1\sigma$ error bands (barely visible, except for the mass bins with fewer halos) are estimated by assuming Poisson counting errors. \textit{Left}: unscaled Betti curves. The curves appear to be self-similar with roughly the same shape, but vastly different physical scales. \textit{Right}: Betti curves scaled by $\ell$ to remove the effect of different number densities.}
    \label{fig:halos_bc}
\end{figure}

\section{Results}\label{sec:results}
Here we present three main results. (1) Betti curves provide complementary information about the distribution of DM haloes compared to that present in the 2-point correlation functions. (2) Galaxies are topologically different from the haloes on small scales. (3) The topology of quiescent galaxies is significantly different from that of the star-forming ones and is especially sensitive to the feedback model used.

\begin{table*}
    \centering
    \begin{tabular}{ccccc}
    \hline
    $\log(M_{200c} / M_{\sun})$ & $N_{\text{haloes}}$ & $\ell$ [$h^{-1}$ Mpc] & $r_0$ [$h^{-1}$ Mpc] & $\gamma$ \\ 
    \hline
    (10.5, 11] & $3.2\times10^5$ & 2.97 & $2.34\pm0.02$ & $-1.36\pm0.01$ \\
    (11, 11.5] & $1.3\times10^5$ & 4.01 & $2.62\pm0.03$ & $-1.46\pm0.01$ \\
    (11.5, 12] & $5.3\times10^4$ & 5.47 & $3.09\pm0.03$ & $-1.53\pm0.02$ \\
    (12, 12.5] & $1.8\times10^4$ & 7.76 & $3.89\pm0.05$ & $-1.69\pm0.03$ \\
    \hline
    \end{tabular}
    \caption{Summary of the TNG300 DM haloes divided into four mass bins to compare their topologies. We list the number of haloes in each bin, the mean particle separation, and the parameters used to fit a power law to the correlation functions.}
    \label{tab:halo_samples}
\end{table*}

\subsection{DM haloes}
In this section, we quantify the topology of the DM halo distribution \changeone{and make comparisons with the log-normal samples to probe the sensitivity of Betti curves to higher order moments of clustering}. Here we focus on the TNG300 simulation due to its large size. We define haloes using a spherical overdensity of 200 times the critical density of the universe, thus all halo masses given refer to $M_{200c}$. We consider haloes above a minimum mass cut of $M_{200c} > 3.2\times10^{10} h^{-1} M_{\sun}$, very roughly corresponding to at least 450 simulation particles in each halo. There are a total of $5.37\times10^5$ such haloes in TNG300, allowing us to further divide them into several mass bins to look at the dependence of topology on halo mass.

We divide our total halo sample into four mass bins based on the value of $\log_{10}(M_{200c}\, [h^{-1}M_{\sun}])$: 10.5 to 11, 11 to 11.5, 11.5 to 12, and \changeone{12 to 12.5}. For each halo sample, we calculate its 2-point correlation function $\xi(r)$ using the Landy-Szalay estimator \citep{Landy1993} implemented in \verb|Corrfunc|\footnote{\url{https://github.com/manodeep/Corrfunc}} \citep{Sinha2021}. \changeone{We estimate the error in the correlation function measurements by dividing the full TNG300 box into 8 equal-size sub-boxes and calculate the standard deviation of the correlation functions calculated from each sub-box.} We then fit a power-law (\autoref{eq:xi}) to each halo sample in order to estimate each sample's clustering length $r_0$. Additionally, we calculate the mean separation $\ell = L / N^{1/3}$ for each sample. The properties of each halo sample are summarized in \autoref{tab:halo_samples}. As expected, we see that higher mass haloes are much rarer and more clustered than lower mass haloes. 

We construct periodic alpha-complex filtrations for each halo sample and compute the corresponding Betti curves (ignoring the infinite persistence intervals as mentioned in \autoref{sec:methods}). \changeone{We also estimate the noise present in the Betti curves by assuming Poisson errors\footnote{Ideally, we would use multiple realizations of the simulation box to estimate the sample variance of the Betti curves. Here, we only have access to the single TNG300 box.}. Since a Betti curve $\beta_d$ represents a number density of topological features inside a volume $L^3$, the sample error should scale roughly as $\sqrt{\beta_d L^3} / L^3 = \sqrt{\beta_d / V}$. The sample error for a scaled Betti curve should scale similarly, but re-scaled by a volume of $\ell^3$. Using numerical experiments, we confirmed this scaling for the case of a Poisson point process.}

In the left column of \autoref{fig:halos_bc}, we plot the unscaled Betti curves normalized by the volume of the simulation box. In these panels, we see that the Betti curves across the different samples have roughly the same shape, but lie at vastly different scales. As seen in \autoref{sec:rand}, this is due to the vastly different number of haloes in each sample: fewer haloes will directly lead to fewer loops or voids present, and those present will occur at larger physical scales. The presence of different clustering lengths $r_0$ also complicates things, but this is a second-order effect. We must first scale out the dependence on the number density of haloes. The right column of \autoref{fig:halos_bc} shows the same Betti curves for the TNG300 haloes, but now re-scaled by $\ell$ and re-normalized to represent the number of homology features within a cube of size $\ell$. \changeone{The peak locations of the Betti curves are all roughly lined up, now that the effect of the different number densities is scaled out.} \changeone{Visually, the scaled Betti curves are very similar across all halo masses, but due to the very small statistical errors that scale as the inverse square root of the number of halos, the small differences are statistically significant. The maximum difference between the Betti curves of adjacent mass bins reaches a nominal value of $14\sigma$ over the considered spatial scales.} In contrast, \cite{Bermejo2022} scaled the Betti curves by the clustering length $r_0$ instead of the mean separation $\ell$ and \changeone{attributed the shift in the locations of the Betti curve peaks to halo topological bias. However, we find that differences between halo mass bins are largely driven by number density, clearly analogous to the random points in \autoref{fig:poisson_betti}.}

\begin{figure}
    \centering
    \includegraphics[width=\linewidth]{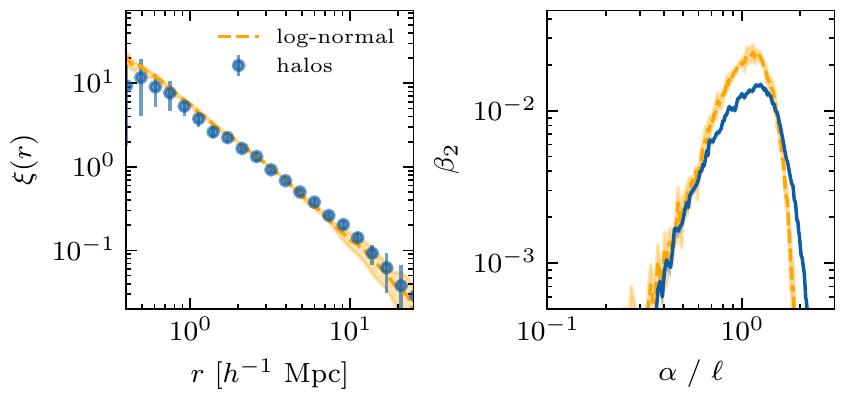}
    \caption{Haloes in the (11.5, 12] mass bin compared to log-normal samples with similar clustering. The \changetwo{left} panel shows the correlation functions (identical by construction), while the \changetwo{right} panel shows the scaled Betti curves $\beta_2$.}
    \label{fig:halo_xi_betti}
\end{figure}

\changeone{Additionally, by comparing the halo Betti curves to the Betti curves of log-normal samples with similar clustering, we see evidence that the Betti curves provide information complementary to that present in the correlation functions. In \autoref{fig:lognormal_betti}, we saw that log-normal fields with higher clustering lengths have Betti curves with smaller amplitudes and longer tails. We see the opposite trend in the halo samples. More massive haloes with higher clustering lengths have more peaked and narrow Betti curves. We show a more concrete comparison in \autoref{fig:halo_xi_betti}. Here, we selected log-normal samples to exactly match the clustering of haloes in the (11.5, 12] mass bin. While the correlation functions are identical by construction, the Betti curves show significant differences. Specifically, we plot the $\beta_2$ Betti curve in order to make comparisons on roughly the same physical scales. On scales of 0.1 to 3$\ell$, or 0.55 to 16.5 $h^{-1}$ Mpc, the correlation functions have a maximum difference of $1.4\sigma$, while the $\beta_2$ curves on the same scales have a maximum difference of $16\sigma$. Similar to the analysis carried out by \cite{Hong2016} and \cite{Naidoo2020} for networks constructed on top of halo catalogs, this shows that Betti curves are indeed sensitive to higher moments of clustering.}

\begin{figure}
    \centering
    \includegraphics[width=\linewidth]{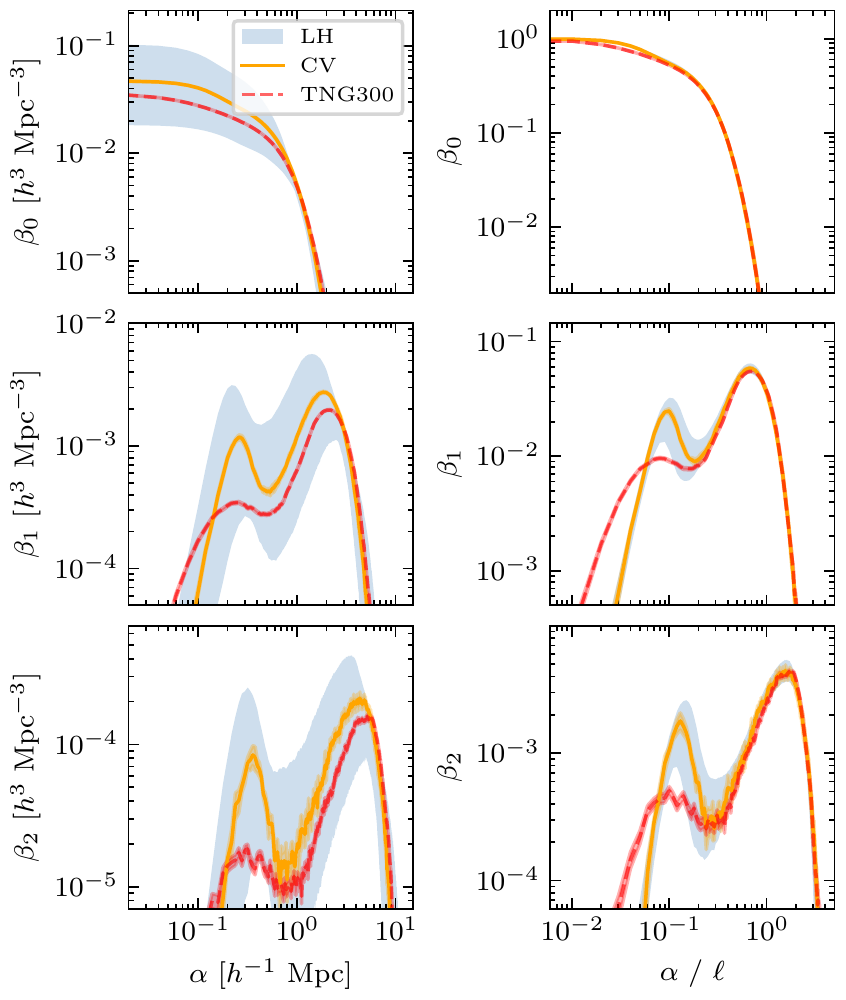}
    \caption{Betti curves for all galaxies in the CAMELS-SAM and TNG300 catalogs. The orange curves (with $1\sigma$ error bands) represent the mean Betti curves for the fiducial CAMELS-SAM model. Similarly, the red curves (with estimated error bands) represent the TNG300 model. The faint blue bands represent the 16 to 84 percentile range of variation of the Betti curves over the full latin hypercube of parameters in the CAMELS-SAM model. \textit{Left}: unscaled Betti curves. \textit{Right}: scaled Betti curves.}
    \label{fig:galaxies_bc}
\end{figure}

\begin{figure}
    \centering
    \includegraphics[width=2in]{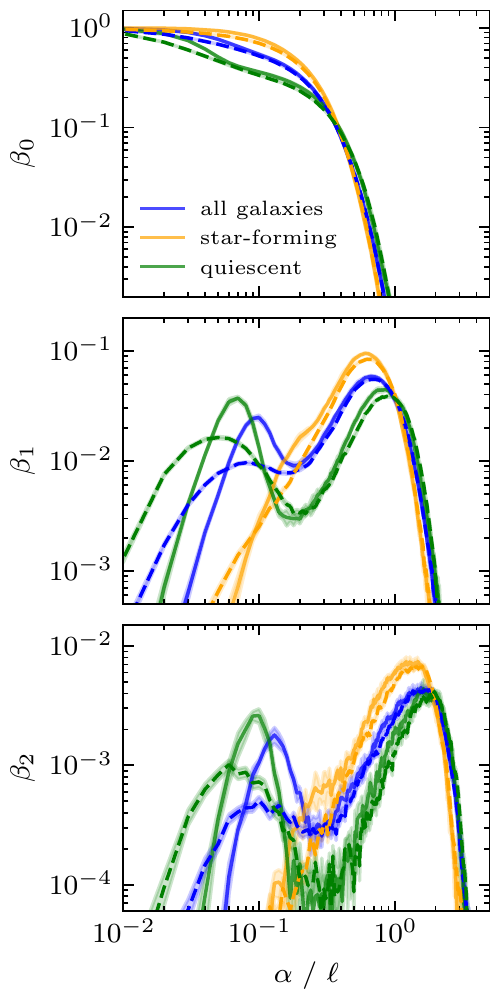}
    \caption{Scaled Betti curves for different galaxy subsets from CAMELS-SAM and TNG300. Here the color of the curve indicates the type of galaxy considered (all galaxies, star-forming, or quiescent), solid lines indicate the CAMELS-SAM fiducial model, and dashed lines indicate the TNG300 model. The bands around the curves roughly indicate $1\sigma$ variance around the mean.}
    \label{fig:galaxies_bc_scaled}
\end{figure}

\begin{figure*}
    \centering
    \includegraphics[width=0.7\linewidth]{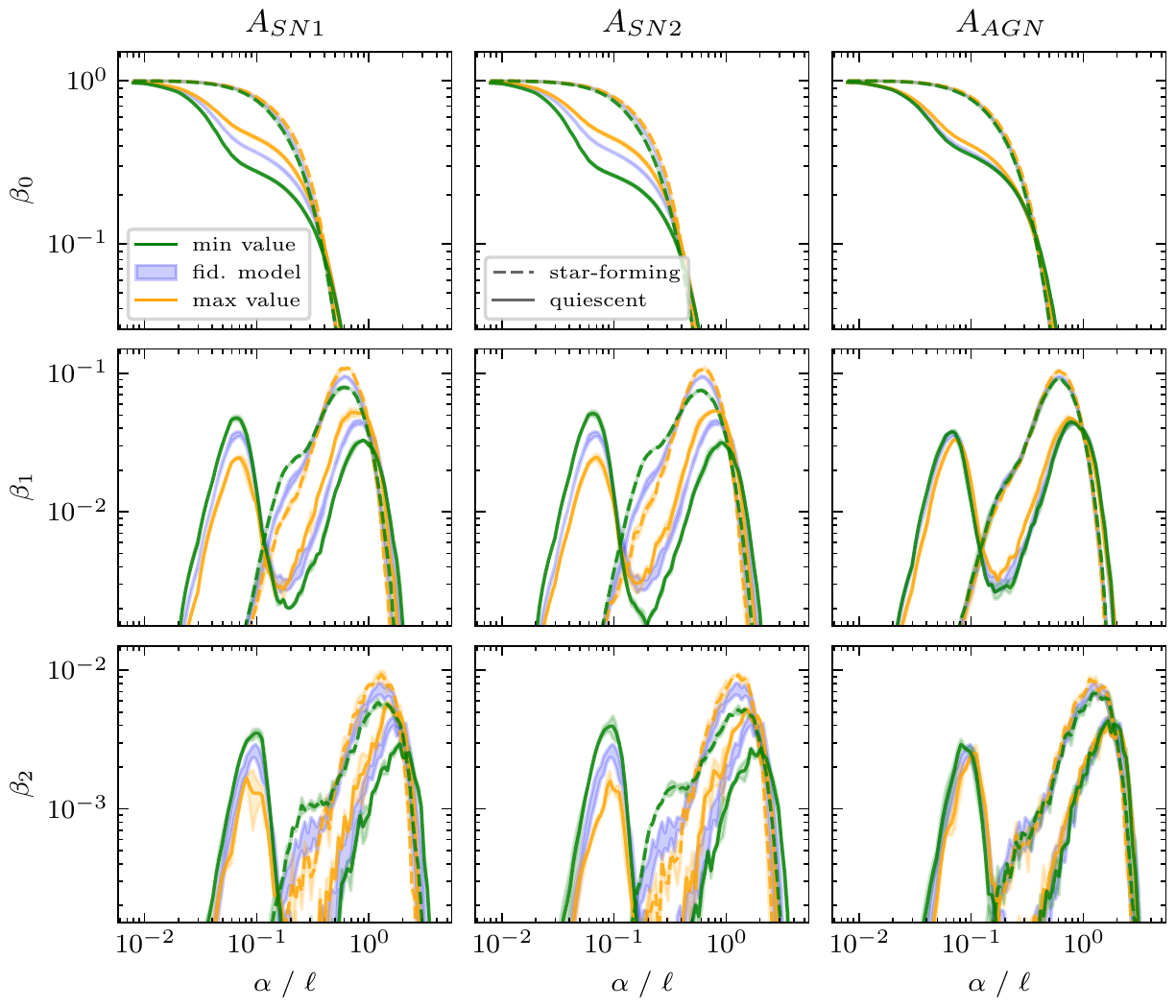}
    \caption{Betti curves for the star-forming and quiescent galaxies from the CAMELS-SAM 1P simulations. Each column shows how variations in the corresponding feedback parameter causes changes in the Betti curves. Here the solid/dashed lines represent the quiescent/star-forming galaxies, while orange/green represent the minimum/maximum value of the corresponding feedback parameter. The faint blue bands in between the lines show the $1\sigma$ variation of the fiducial CAMELS-SAM model.}
    \label{fig:sf_qsnt_1p}
\end{figure*}

\subsection{Galaxies}
Next, we look at galaxies and various sub-selections of galaxies. We use both CAMELS-SAM and TNG300 in order to compare two different models of galaxy formation. In the TNG300 simulation, we select all subhaloes with at least a 10\% DM mass fraction in order to filter out a small population of baryonic lumps, most likely produced by disc fragmentation \citep[as noted by][]{Springel2018}. The CAMELS-SAM catalogs already only include galaxies due to the SC-SAM directly modelling galaxy formation inside the DM haloes. For both simulations, we additionally impose a minimum stellar mass cut of $5\times10^8 h^{-1} M_{\sun}$. We further divide both galaxy samples into star-forming and quiescent galaxies. Following \cite{EuclidCollaboration2023}, we define the boundary between star-forming and quiescent to be a specific star formation rate (sSFR) of $10^{-10.5}$ yr$^{-1}$ = 0.0316 Gyr$^{-1}$. 

\autoref{fig:galaxies_bc} shows the Betti curves for all of the galaxies from the CAMELS-SAM and TNG300 samples. \changeone{The two columns again show that scaling the Betti curves by $\ell$ removes the effect of very different galaxy number densities in the different simulations, while in the CAMELS-SAM LH set there is still significant variance remaining due to the variation of cosmological and astrophysical parameters. We also see that these Betti curves capture the small-scale clustering of galaxies inside clusters. Both the $\beta_1$ and the $\beta_2$ Betti curves have a second peak at small scales that was not present in the halo Betti curves or the random samples in \autoref{sec:rand}. This peak, occuring at a physical scale of $\sim 200h^{-1}$ kpc, directly corresponds to structure inside galaxy clusters. We confirmed this by excluding from our analysis all galaxies inside very massive halos. The small-scale peak decreases in amplitude when all galaxies in halos more massive than $10^{14}h^{-1}$ $M_{\sun}$ are excluded and is practically gone once we exclude all galaxies in halos $> 10^{13}h^{-1}$ $M_{\sun}$. We note that this is similar to what was seen by \cite{Pranav2017} in the context of heuristic models of the cosmic web: multiple scales naturally show up in the topological summaries of hierarchical clustering models.} 

\changeone{Additionally, these Betti curves reveal differences between different models of galaxy formation. At large spatial scales ($> 1\ell$) the Betti curves for TNG300 and the fiducial CAMELS-SAM model are roughly consistent, with the $\beta_0$ and $\beta_1$ curves attaining at most a $2\sigma$ difference between the TNG300 and CAMELS-SAM models. The $\beta_2$ curves attain a maximum difference of roughly $3.5\sigma$ on the same scales. On small scales, the Betti curves for the TNG300 and SC-SAM models are significantly different, due to inherent differences in the galaxy formation models. All three Betti curves reach maximum differences of over $10\sigma$ when comparing the TNG300 and fiducial CAMELS-SAM models at scales $< 1\ell$.}

\autoref{fig:galaxies_bc_scaled} shows the scaled Betti curves for all three samples of galaxies (all, star-forming, quiescent) for both TNG300 and the fiducial CAMELS-SAM model. Here we see that there are significant differences between the Betti curves of quiescent and star-forming galaxies. The peak at small scales only appears in the curves for the quiescent galaxies. This supports our \changeone{conclusion} that the small scale peak is due to galaxy clusters, as galaxies in large clusters tend to be quiescent \cite[see, for example,][]{Peng2010,Wetzel2013}.

\subsection{Dependence on feedback parameters}
Finally, we use the CAMELS-SAM 1P set of simulations to test the effect of various feedback parameters on the topology of galaxy distributions. We have three parameters that are varied: $A_{\text{SN1}}$, $A_{\text{SN2}}$, and $A_{\text{AGN}}$. For each parameter, there are two simulations (with different initial seeds) that use its maximum value and two that use its minimum value. All other parameters are kept at their fiducial values.

As before, for each simulation we select samples of star-forming and quiescent galaxies and compute their scaled Betti curves. We additionally average each pair of Betti curves to get mean Betti curves for the maximum and the minimum values of each feedback parameter. 

In \autoref{fig:sf_qsnt_1p} we show how the Betti curves of star-forming and quiescent galaxies vary in response to these three parameters. The quiescent galaxies are much more sensitive to changes in these parameters than the star-forming galaxies. The strength of the AGN feedback has comparatively little effect on the topology of these galaxy samples, while the supernova parameters have a very significant effect on the topology of the quiescent galaxies. Increasing the strength of the supernova feedback increases the number of $H_1$ and $H_2$ features at large scales and suppresses the peak at small scales. \changeone{We show in \autoref{tab:gal_diffs} the largest differences between Betti curves and the fiducial model compared to the expected variance, with the differences for the SNe parameters being nominally more than $30\sigma$.} This suggests that these Betti curves should be able to differentiate different models of feedback. \changeone{We plan to do a more detailed analysis to determine how well different summary statistics can constrain these parameters in future work.}

\begin{table}
    \centering
    \begin{tabular}{lccc}
    \hline
     & $A_{\text{SN1}}$ & $A_{\text{SN2}}$ & $A_{\text{AGN}}$ \\ 
    \hline
    $\beta_0$ & 33 (16) / 15 (19) & 37 (29) / 14 (16) & 8.0 (7.4) / 22 (6.6) \\
    $\beta_1$ & 39 (14) / 27 (15) & 38 (17) / 17 (6.1) & 24 (3.8) / 8.4 (9.3) \\
    $\beta_2$ & 5.6 (7.9) / 4.1 (4.8) & 6.9 (10) / 6.0 (4.3) & 3.7 (2.3) / 2.7 (2.8) \\
    \hline
    \end{tabular}
    \caption{Maximum differences (in $\sigma$'s) between the fiducial model Betti curves and the Betti curves for which one of the feedback parameters is at an extreme value. For each feedback parameter and Betti curve we report the maximum difference between the Betti curves for quiescent (star-forming) galaxies in the fiducial model and in the model with the minimum / maximum value of the parameter.}
    \label{tab:gal_diffs}
\end{table}

\section{Summary and future outlook}\label{sec:conlusion}
We used Betti curves from persistent homology to characterize the large-scale structure of simulated galaxies and DM haloes. To our knowledge, this is the first time such techniques from TDA have been applied to catalogs of both galaxies and haloes.

We showed that these topological summaries reveal differences in the large-scale structure of haloes and different galaxy populations. First, we \changeone{showed that the number density of points is the main factor that determines the scale of the resulting Betti curves, but we are able to compare across point samples with different number densities after scaling by the mean interparticle spacing $\ell$. Using DM haloes, we showed that Betti curves are sensitive to higher order moments of clustering beyond the 2-point correlation function. The Betti curves for galaxies are very sensitive to small-scale clustering;} we found that a second physical scale shows up in the topology of galaxies that is not present in the haloes. We identify this additional structure present at small scales as coming from galaxies inside clusters. This is supported by the fact that this peak in the Betti curves is present only in the quiescent galaxies; the star-forming ones look very similar to the DM haloes. Further, by looking at parameter variations in the CAMELS-SAM simulations, we found that the topology of quiescent galaxies is especially sensitive to differences in the feedback model used. Based on these results, we argue that galaxies are topologically distinct from DM haloes and that topological summaries could provide very useful information in order to better constrain models of galaxy formation and evolution.

This paper also adds to the body of literature advocating for the use of summary statistics that probe higher-order aspects of clustering.  Persistent homology provides a very natural way of quantifying the topology of large-scale structure in the universe, due to the wide range of physical scales of interest.

There are many future directions for this work. So far we have only looked at the Betti curves for the final $z=0$ snapshots of simulations. It would be very interesting to look at the redshift evolution of the Betti curves and to construct lightcones to investigate the topology of galaxies in redshift space. Potentially, looking at the time evolution of the redshift-space topology of quiescent galaxies could help constrain when and how star formation shuts off inside clusters. 

Additionally, there are many more simulated catalogs that this method could be applied to in order to check if these topological features are consistent across all models. Large hydrodynamic cosmological simulations are constantly improving (MillenniumTNG \cite{Pakmor2022}, in particular, looks extremely promising) and we hope to fully utilize them to better understand the constraining power of these topological summaries. We are also interested in comparing alternatives to full hydrodynamic simulations, for example other semi-analytic models or halo occupation distribution models.

In the end, we hope to apply these techniques to observational data. We hypothesize that topological summaries, like the Betti curves used here, could provide useful observational constraints to compare to simulations of galaxy formation.

\section*{Acknowledgements}
We thank the anonymous referee for their very useful comments that helped to improve this work.

We thank the TNG and CAMELS collaborations for publicly releasing their simulation data and example analysis scripts.

The analysis in this work was mainly carried out using Python. In addition to the previously cited software, the following packages were extremely useful: numpy \citep{Harris2020}, scipy \citep{Virtanen2020}, matplotlib \citep{Hunter2007}, IPython \citep{Perez2007}, h5py, and mpi4py \citep{Dalcin2021}. 

This work made use of the Illinois Campus Cluster, a computing resource that is operated by the Illinois Campus Cluster Program (ICCP) in conjunction with the National Center for Supercomputing Applications (NCSA) and which is supported by funds from the University of Illinois at Urbana-Champaign.

\section*{Data Availability}
All simulation data used in this work is publicly available.
The analysis code has been posted on Github\footnote{\url{https://github.com/ajouellette/galaxy-topology}}.
Any other data will be shared upon reasonable request to the corresponding author.



\bibliographystyle{mnras}
\bibliography{references}

\bsp	
\label{lastpage}
\end{document}